\def\rn{\noindent\parshape 2 0truecm 8.8truecm 0.3truecm 8.5truecm}
\def\nn#1 #2{#1, #2.}				% Name with 1 initial
\def\nnn#1 #2 #3{#1, #2. #3.}			% Name with 2 initials
\def\nnnn#1 #2 #3 #4{#1, #2. #3. #4.}		% Name with 3 initials
\def\nnnnn#1 #2 #3 #4 #5{#1, #2. #3. #4. #5.}	% Name with 4 initials

\def\rg#1;#2;#3;#4;#5;#6 {\par\rn#1 #2, {\it #3}, {\bf #4}, #5 (``#6'') \par}
\def\rf#1;#2;#3;#4;#5 {\par\rn#1 #2, {\it #3}, {\bf #4}, #5\par}
\def\rfbook#1;#2;#3;#4;#5 {{\frenchspacing\par\rn#1 #2, {\it #3} (#4: #5)\par}}
\def\rfproc#1;#2;#3;#4;#5;#6 {{\frenchspacing\par\rn#1 #2, in {\it #3}, ed. #4 (#5: #6)\par}}
\def\rfprep#1;#2;#3  {{\par\rn#1 #2, #3\par}}
\def\rfprepp#1;#2;#3 {{\par\rn#1 #2, #3\par}}

\def\Mpc{{\rm Mpc}}

\def\km{{\rm km}}
\def\s{{\rm s}}

\def\beq#1{\begin{equation}\label{#1}}
\def\eeq{\end{equation}}
\def\beqa#1{\begin{eqnarray}\label{#1}}
\def\eeqa{\end{eqnarray}}

\def\spose#1{\hbox to 0pt{#1\hss}}
\def\simlt{\mathrel{\spose{\lower 3pt\hbox{$\mathchar"218$}}
     \raise 2.0pt\hbox{$\mathchar"13C$}}}
\def\simgt{\mathrel{\spose{\lower 3pt\hbox{$\mathchar"218$}}
     \raise 2.0pt\hbox{$\mathchar"13E$}}}
\def\simpropto{\mathrel{\spose{\lower 3pt\hbox{$\mathchar"218$}}
     \raise 2.0pt\hbox{$\propto$}}}

\def\ed{\end{document}}

\def\L{{\cal L}}

\def\k{{\bf k}}
\def\p{{\bf p}}
\def\d{{\bf d}}

\documentstyle[emulateapj,danonecolfloat]{article}
\begin{document}
\twocolumn[%%% Begin front material

%%%%%%%%%%%%%%%%%%%%%%%%%%%%%

%\tighten
%\eqsecnum
%\received{4 August 1988}
%\accepted{23 September 1988}
\journalid{337}{15 January 1989}
\articleid{11}{14}

%\submitted{Submitted to ApJ May 20, 1999}
\submitted{\today. To be submitted to ApJ.}
%\submitted{Submitted to ApJL September 16; accepted February 2}

\title{Constraints from the Lyman $\alpha$ forest power spectrum}
\author{Matias Zaldarriaga}
\affil{Institute for Advanced Study, School of Natural Sciences,
Olden Lane, Princeton, NJ 08540}

\vskip 1pc

\author{Lam Hui}
\affil{Department of Physics, Columbia University, New York, NY 10027\\
Institute for Advanced
Study, School of Natural Sciences, Olden Lane, Princeton, NJ 08540}

\vskip 1pc

\author{Max Tegmark}
\affil{Department of Physics, University of Pennsylvania,
Philadelphia, PA 19104}

\vskip 1pc


\begin{abstract}
We use published measurements of the transmission power spectrum of
the Lyman $\alpha$ forest to constrain several parameters that
describe cosmology and thermal properties of the intergalactic medium
(IGM).  A 6 parameter grid is constructed using Particle-Mesh dark
matter simulations together with scaling relations to make predictions
for the gas properties. We fit for all parameters simultaneously and
identify several degeneracies. We find that the temperature of the IGM
can be well determined from the fall-off of the power spectrum at
small scales. We find a temperature around $2 \times 10^4$ K,
dependent on the slope of the gas equation of state. We see no
evidence for evolution in the IGM temperature.  We place constraints
on the amplitude of the dark matter fluctuations. However, contrary to
previous results, the slope of the dark matter power spectrum is
poorly constrained.  This is due to uncertainty in the effective Jeans
smoothing scale, which depends on the temperature as well as the
thermal history of the gas.
\end{abstract}


]%%% End front material

%%%%%%%%%%%%%%%%%%%%%%%%%%%%%%%%%%%%%%%%%%%%%%

\section{Introduction}

The past decade has seen remarkable progress in our understanding
of the Lyman $\alpha$ forest. Comparison of the absorption line data 
and numerical simulations has lead to 
a clear picture for the forest sometimes called the fluctuating
Gunn Peterson effect
(\cite{chen94,her95,zhang95,m96,mucket96,wb96,th98}). In this picture
most of the absorption is produced by low density unshocked gas in the
voids or mildly overdense regions in the universe.  This gas is in
ionization equilibrium and traces broadly the distribution of the dark
matter, but is also sensitive to its equation of state. Simple
semi-analytic models based on these ideas have been developed and
shown to be successful in explaining the main features of numerical
simulations (\cite{Bi92,rm95,BD97,GH96,croft97,HuiGne97,Hui97b}).

The main ingredient that determines the absorption in the forest is
the distribution of the dark matter so the forest can be a very
powerful probe of cosmology. The probability distribution of the transmission
has also been computed and successfully compared to the
data (\cite{rauch97,nh99,McD99}).  It has been shown that the moments of this
distribution can be predicted analytically using the known scalings
for the matter and simple ideas of local biasing (\cite{gaztacroft}).

Perhaps the most important application of the forest is to measure the
power spectrum of the dark matter at redshifts around $z\sim 3$, which
can place strong constraints on cosmology and the nature of
dark matter (\cite{croft98,croft99,wc2000,naray00}). 
Croft et al. (1998) pioneered the approach of inverting 
the shape of the mass power spectrum directly from the 
1-D power spectrum of the flux or transmission. 
The amplitude of the mass power spectrum is then obtained by
comparing with simulations.
Hui (1999) pointed out the possibility of a bias
in the recovered shape due to redshift distortions, and
suggested a modified inversion technique (see also \cite{Mcdjordi}).
Most work so far focused on cosmological information that
could be obtained from the large scale forest power spectrum.

In this paper we obtain cosmological information from
the transmission power spectrum of the
forest using a different approach. Rather than trying to extract the
power spectrum of the dark matter by 
applying some inversion technique to the data,
we take the observed transmission power spectrum as is, and
simply compare it with predictions from a wide range of models.
This is in part the approach taken by McDonald et al. (1999), but
the type of models they examined had a fixed thermal evolution. 
Here, we construct a grid of models described
by 6 parameters and constrain all parameters simultaneously
using a likelihood analysis analogous to what has been developed for 
cosmic microwave background (CMB)
data (eg. \cite{tz00}). 

Our analysis allows us to use the information on smaller length
scales. In particular, the sharp drop in the power spectrum of the
forest on small scales (see figure \ref{bestfitdata}) is a direct
consequence of thermal broadening. The fact that the small scale power
could be sensitive to the temperature has been noted in the past
(eg. \cite{th00}). We are able to place constraints on the temperature
of the IGM at high redshift.

There has been some controversy in the literature as to what the
equation of state of the intergalactic medium (IGM) is as determined by 
the width of
absorption lines. Different studies have reached different conclusions
about the time evolution of the gas temperature and its equation of
state (\cite{joop99,McD00}). Our results can be directly compared to 
those obtained by these other techniques.

The outline of the remainder of the paper is as follows: 
we describe the data used in our analysis in section
\ref{data}, describe our method in \ref{method}, present
constraints on parameters in \ref{results}, suggest ways to
improve our analysis in the future in \ref{future} and conclude 
in \ref{conclusions}.

\section{Data}\label{data}

\begin{figure}[tb]
\centerline{\epsfxsize=9cm\epsffile{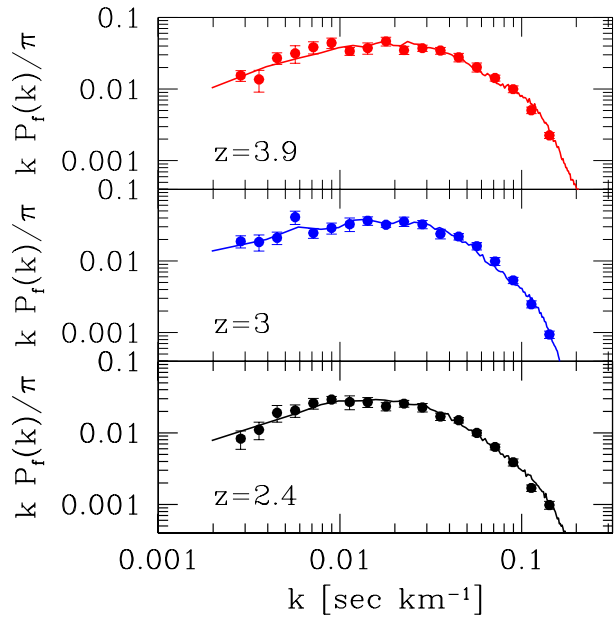}}
\caption{Transmission power spectrum from from McDonald et al. (1999)
together with our best fits.}
\label{bestfitdata}
\end{figure}

The most detailed measurements of the transmission power spectrum of the
forest in the literature have been presented by McDonald et al (1999). We will
use these data for our analysis. Figure \ref{bestfitdata} shows the
data together with the best fit models in our grid. The data points
and error bars were taken straight from the tables in McDonald et al. (1999),
except
for the last two points in the plot which were only used as upper
limits because of concerns about metal line contamination.

In addition, we used the mean transmission as determined by McDonald et al. (1999)
with their quoted errors (table \ref{meanFtab}).
\begin{table}[ht]
\center{
\begin{tabular}{||c|c|c||}\hline
$\bar{z}$ & $\langle F \rangle$ & $\sigma^2_F$ \\
\hline

3.89 & $0.475\pm0.021$ & $0.1293\pm 0.0030$ \\
3.00 & $0.684\pm0.023$ & $0.1174\pm 0.0056$ \\
2.41 & $0.818\pm0.012$ & $0.0789\pm 0.0068$ \\
\hline
\end{tabular}}
\caption{Mean ($\bar F = \langle F \rangle$) and variance ($\sigma^2_F$) of the 
transmission $F = e^{-\tau}$ as measured
by \cite{McD99} for several redshift bins.}
\label{meanFtab}
\end{table}

\section{Method}\label{method}

\subsection{Model Predictions}

The model predictions were done using the PM model for the forest
(\cite{croft98,mw00}).  In
this model PM simulations are run to compute dark matter densities and
velocities. The gas density and temperature are calculated
using simple scaling relations inspired by the results of full
hydrodynamics simulations. Our objective is to constrain the
parameters of these scaling relations.

We ran 7 PM simulations using $128^3$ particles with a box size $16
h^{-1} \Mpc$ (corresponding to $3200 {\rm km/sec}$ at redshift $z=3$)
of the standard Cold Dark Matter (SCDM) model with different spectral
indices $n=(0.4, 0.7, 0.9, 1, 1.1, 1.3,$ $1.7)$ for the power spectrum
of initial density perturbations.  Having a non-vanishing cosmological
constant will not significantly alter our results because $\Omega_m$
is close to $1$ at the relevant redshifts regardless of whether
$\Lambda = 0$ -- the main determining factor is instead the power
spectrum at certain scales, which we will constrain.  We stored the
density and velocity at expansion factors $a=\left(0.05, 0.065, 0.085,
0.11, 0.14, 0.19, 0.24, 0.31, 0.41 \right)$. The expansion factors are
used to span a range of power spectrum normalizations.

The baryon density is obtained by smoothing the dark matter density;
the smoothing mimics the effect of pressure forces.  Following Gnedin
\& Hui (1998) we adopt $\delta_b (\k)=W_f(k) \delta_{CDM}(\k)$ where
$W(k)$ is a Fourier space smoothing, which we take to be
$W_f(k)=\exp(-(k/k_f)^2)$.  As discussed in Gnedin \& Hui (1998),
$k_f$ is proportional to the Jeans scale but the constant of
proportionality depends on the details of the reionization history of
the universe (see also Nusser 2000). For this reason, we treat $k_f$
as a free parameter.  At redshift $z\sim 3$, the filtering is expected
to be around $k_f\sim 35 \ {\rm h\, Mpc^{-1}}$ (Gnedin \& Hui 1998).

The optical depth for each gas element is related to the overdensity
$\tau$ using a power law (\cite{HuiGne97}), 
\beq{taudelta} 
\tau=a_0 \Delta^{\beta} 
\label{taua0}
\eeq
where $\Delta=\rho/\bar \rho$ denotes the gas overdensity.  The transmission
is $e^{-\tau}$. The
constant $a_0$ is fixed so that the generated spectra have a
particular value of the mean transmission (which we will call
$\bar f$). 
We make a grid of mean transmissions but use the observed
measurements (in table \ref{meanFtab}) as a prior in our likelihood.
Our grid of mean transmissions serves effectively as a grid
of $a_0$'s.

For each grid point in the box a temperature is assigned to the gas 
using a simple power law equation of state for the gas, 
\beq{t0delta}
T=T_0 \Delta^{\alpha}. 
\eeq
We treat $T_0$ and $\alpha$ as free parameters. The power law index
$\beta$ in equation (\ref{taudelta}) is related to $\alpha$,
$\beta=2-0.7\alpha$. Smoothing due to thermal motions is included when
the spectra are generated. The absorption produced by each fluid
element is distributed in velocity space as $\exp(-(\Delta
s/b)^2)/b\sqrt{\pi}$, where $\Delta s$ is the velocity space
separation and the $b$ parameter is given by $b=\sqrt{2kT/m_p}$
$\approx 13\ \km \ \s^{-1} $ for $T = 10^4$ K.

Our parameter vector $\p=(a,n,k_f,T_0,\alpha,\bar f)$ has 6 dimensions. 
We created a grid of model predictions for each choice of parameters
in a grid,
\begin{itemize}
\item $a=(.05, .065, .085, .11, .14,
.19, .24, .31, .41)$
\item $n=(0.4, 0.7, 0.9, 1, 1.1, 1.3, 1.7)$
\item 
$k_f=(5, 10, 20, 30, 40, 45, 50, 55, 60, 70, 80)$
\item $T_0=\left (150, 200, 250, 300, 350, 400, 450, 500, 550, 600, \right.$ 
 $\left. 700 \right)$
\item $\alpha=(0.0. 0.1, 0.2, 0.3, 0.4, 0.5, 0.6)$
\item $\bar f=\left (.4, .45, .475, .5, .525, .55, .575, .6, .625, .65, .665,
\right.$ $\left. .684, .7, .725, .75, .78, .8, .82, .85, .9 \right)$
\end{itemize}
where by convention the scale factor is $a=1/(1+z)$, $k_f$ is measured
in ${\rm h Mpc^{-1}}$ and $T_0$ in $(\km \s^{-1})^2$. Thus we have a
total of $9 \times \ 7 \times \ 11 \times \ 11 \times \ 7 \times 20 =
1067220$ combinations of model parameters. For each point in our grid
we extract 500 lines of sight randomly from the simulations, generate
spectra and measure the power spectra of the forest. We measure the
power spectra of the relative transmission fluctuations, $\delta F/ \bar F$ where
$F =\exp(-\tau)$. 

\subsection{Likelihood calculation}

For each model in our grid, we compute the likelihood by comparing the
power spectra computed form the simulations to the points in McDonald
et al. (1999).  In this paper, we will stick to a crude Gaussian
approximation, 
\beq{chi2eq} 
\L(\d;\p)\propto  \prod_i
\exp\left[-{1\over 2}\left({d_i-P_{fi}(\p) \over
\sigma_i}\right)^2\right], 
\label{like1}
\eeq 
where $i$ runs over the different data points $d_i$, the model
prediction for that wavevector are $P_{fi}(\p)$ and $\sigma_i$ are the
error bars on each point.  By comparing different realizations we
computed the statistical errors in our model predictions, which were
approximately $5\%$. We added those in quadrature to the observational
errors to compute $\sigma_i$. The addition of this extra source of
variance makes very little difference because the observational errors
are significantly larger. Most certainly systematic errors associated
with our simplified model of the forest will dominate the error
budget. We also used the last two points ($k \ge 0.1 {\,\rm km^{-1} \, \s} $) as
upper limits. To do so in practice we increased $\sigma$ for those 
points to $\sigma=0.3 P_{f}$, roughly a factor of three increase on
the quoted error bars and used it as a one sided error bar.  For models
that had more power than $P_{f}$ we computed $\chi^2$ with the
increased error, for models with less power, we set the contribution
to $\chi^2$ from that point to zero.

The data points reported in McDonald et al. (1999) are for the power
spectrum of the transmission $\delta F$, and this is what
we directly compare our predictions with (see, however, \S \ref{future}
for virtues of using power spectrum of $\delta F/\bar F$ instead).
We furthermore add to the
likelihood a term to account for the measurement of $\bar F$. We
multiply the likelihood in \ref{like1} by $\exp(-(\bar f -\bar
F)^2/\sigma_{F}^2)$ where $\bar f$ is the parameter in our grid and $\bar
F$ is the observed value at the appropriate redshift.  The error in
the mean transmission is denoted $\sigma_{F}$.

The full likelihood function is $\L=e^{-\chi^2/2}$, where $\chi^2$ is
simply the chi-squared goodness of fit of the model to the data. We
have chosen to keep things this simple because we are in any case
unable to eliminate a major source of inaccuracy: most probably there
are correlations between the estimates of the power on different
scales which are hard to estimate. We are using 18 bins for the power
spectrum, thus estimating a covariance matrix requires estimating 153
numbers. Naive estimates of the covariance matrix based on a limited
number of lines of sight are very noisy which translates into
artificially low probabilities for some models, thus they cannot be
used in this analysis. More theoretical work needs to be done in order
to address this problem in a satisfactory manner. Perhaps one could
look for guidance in simulations to build a simple parametrised model
of the covariance matrix and then use the scatter between the lines of
sight to fix those parameters.  Moreover the window function needed to
relate observed and theoretical power spectra are not available for
this observation. We choose to use flat windows inside each bin but
also tried Gaussian windows, which made only minor differences.  There
is ample room for improvement in this part of the calculation.

Once we have computed the likelihood for every model in our grid we
marginalize along one direction at a time until we get one- or
two-dimensional constraints on parameters. We use the method developed
in Tegmark \& Zaldarriaga (2000) for this purpose.

\section{Results}\label{results}

In this section we report the constraints we have obtained.  We
first discuss our results on the equation of state of the IGM, then on
the relation between baryons and dark matter and finally we summarize
our constraints on the power spectrum of primordial fluctuations.

\subsection{The temperature of the IGM}

\begin{figure}[tb]
\centerline{\epsfxsize=9cm\epsffile{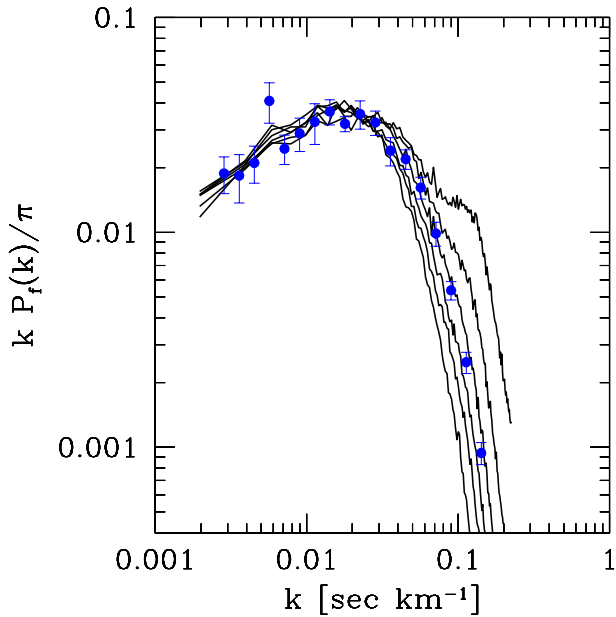}}
\caption{Transmission power spectrum at $z=3$ for models with varying
$T_0$ ($T_0 = (150,250,\cdots,650) (\km/\s)^2$ increasing from top to
bottom). All other parameters are left constant, $\p=(.24, .7 , 50,
T_0, .2,.7)$. The data points are from McDonald et al. (1999).}
\label{movieps}
\end{figure}

The shape of the power spectrum of the forest has a sharp cut-off
around $k=0.02\ {\rm km^{-1}  \ \s}$ which is due to the thermal
broadening of the lines. Figure \ref{movieps} shows a sequence of
models with increasing IGM temperature. As the temperature increases
the smoothing becomes more important and the power on small scales is
reduced. At $k\sim 0.04\ {\rm km^{-1}\ sec}$ the power is down a factor
of 2 from the extrapolation of the larger wavelengths. This
corresponds to a temperature $T\sim 1/k\sim 2 \times 10^{4} K$.  Our
likelihood analysis takes advantage of this dependence to find
constraints on the temperature.

\begin{figure}[tb]
\centerline{\epsfxsize=9cm\epsffile{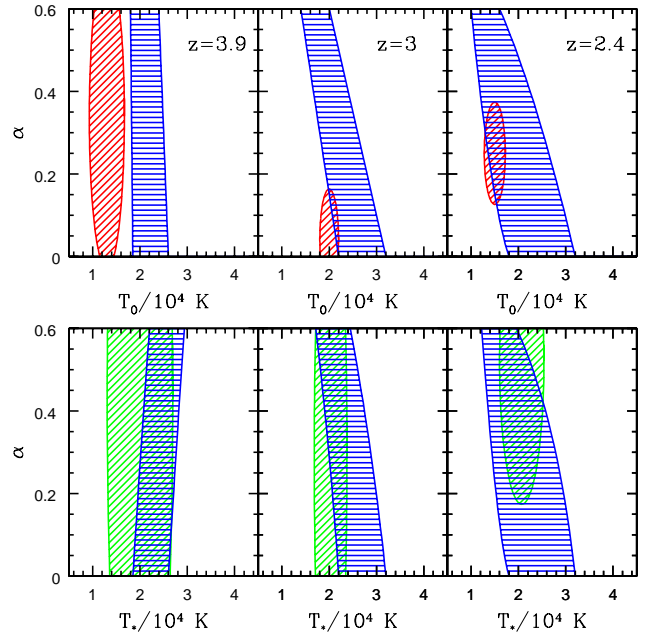}}
\caption{Constraints in the $T-\alpha$ plane (95 \% as described in
the text). The top panel corresponds to the mean density while the
bottom panel is for an overdensity $\Delta_*$. The ellipses with lines
at $45^o$ angle on the top
are the equivalent constraints from Schaye et al. (2000). and on the bottom
from McDonald et al. (1999)}
\label{t0alconst}
\end{figure}

Figure \ref{t0alconst} summarizes the constraints we obtained.  It
shows $95 \%$ confidence regions\footnote{The lines that enclose our
allowed regions are actually lines of constant $\chi^2$. In particular
we use $\Delta \chi^2=6.18$, which would correspond to 95\%
of the area under a 2 dimensional Gaussian.} in the $T_0-\alpha$
plane. Only $\alpha$'s between $0$ and $0.6$ are examined, because
realistic $\alpha$ are expected to fall in this range (\cite{HuiGne97}).
We find a degeneracy in this plane, which is
caused by the fact that our constraint is most tight for the
temperature at a density different from the mean density. Under the
assumption that the temperature density relation is a power law, the
temperature at any given overdensity $\Delta_*$ can be calculated from
$(T_0,\alpha)$ as:
\begin{equation}
T_* = T_0 \ \Delta_*^\alpha.
\label{tstar}
\end{equation}
Our method to determine the temperature is most sensitive to $T_*$, so
models with the same $T_*$ are all good fits to the data. This set of
models correspond to lines in the $T_0-\alpha$ plane.

\begin{figure}[tb]
\centerline{\epsfxsize=9cm\epsffile{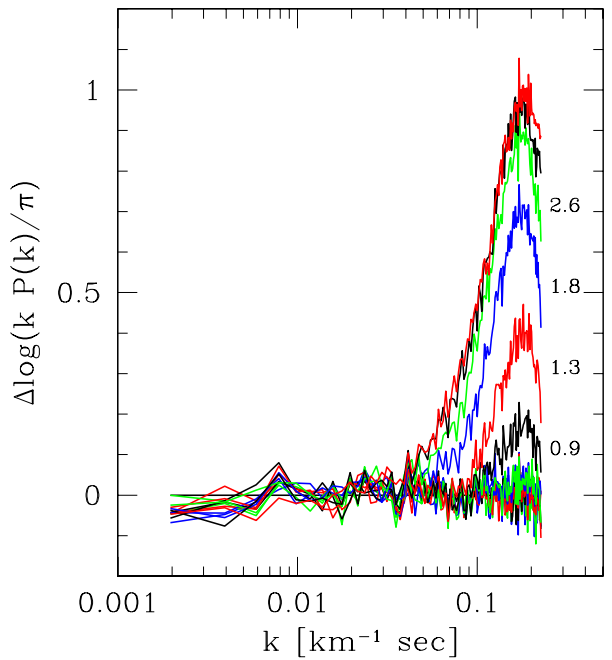}}
\caption{Differences in the power spectrum of the flux of the forest
at redshift $z=3$ for
models with a step like temperature density relation as described in
the test. Labels on the right indicate the value of $\Delta_{jump}$}
\label{overdensity}
\end{figure}

To investigate what $\Delta_*$ is, we run simulations in which
the temperature-density relation has a step at a fixed overdensity
$\Delta_{jump}$, 
\begin{eqnarray}
T &=& T_0\ ({\rm if}\ \Delta < \Delta_{jump}) \nonumber \\  
  &=& 2 T_0\ ({\rm if}\ \Delta > \Delta_{jump}) 
\end{eqnarray}
with $T_0 = 1.2 \ 10^{4} {\rm K}$. In figure \ref{overdensity} we show
the change in the transmission power spectrum as we change the value
of $\Delta_{jump}$ for a model that fits the observations at redshift
$z=3$. For low values there is no change. When $\Delta_{jump}\sim 1$
changes become noticeable but once $\Delta_{jump}> 1.8$ further
increases in $\Delta_{jump}$ no longer change the power spectrum.  At
redshift $z=3$ the transmission power spectrum is therefore sensitive
to $0.9 < \Delta < 1.8$. Given that our method is most sensitive to
$\Delta$'s larger that one, the contours of equal likelihood are
tilted in the $(T_0,\alpha)$ plane in figure \ref{t0alconst}. On the
other hand, because the method is intrinsically sensitive to a range
of overdensities, it should be possible to extract information about
$\alpha$ once the errors in the measured power decrease.  The bottom
panel of figure \ref{t0alconst} shows our constraints at
$\Delta_*=1.4$ constructed by reparametrizing our model grid using
equation \ref{tstar}.

A complementary method to determine the temperature was proposed in
Schaye et al. (1999) and implemented in McDonald et al. (2000),
Ricotti et al. (2000)and Schaye et al. (2000).  The method is based on
the observation that the width of the lines in the spectrum has
several contributions, from thermal broadening, peculiar velocities
and the intrinsic sizes of the structures (\cite{huirut98}). The
minimum width of the lines is thus a measure of the thermal
broadening. In the three papers cited above, the distribution of
widths of lines was used to infer constraints on the temperature, two
of which are shown as ellipses in our plot. We emphasize that those
constraints were obtained fitting lines and determining a minimum
width in their distribution. Our constraint did not require the
fitting of lines. Although the analyses of Schaye et al. (1999) and
McDonald et al. (2000) are based on the same underlying idea and very
similar data sets, they differ in several technical details and agree
neither in their conclusions nor in the size of their error bars.

Our results are in good agreement with those of McDonald et al. (2000)
and we have comparable error bars. Our temperature is higher than that
of Schaye et al. (2000). This could signal a bias in one of the methods, an
underestimate of the errors or some unaccounted for physical
effect. For example, to interpret the line fitting results one needs to
know what gas density is responsible for the majority of the absorption. Simulations
are usually used for this purpose and errors in this determination can
lead to changes in the inferred temperatures. If the difference is not
systematic, then a particularly interesting possibility is an
inhomogeneous equation of state, which could be produced by a recent
reionization of Helium II (see e.g. Madau 2000 for a review). 
In that case, the temperature measured from
the cut-off in the line width distribution would tend to be lower
because the method tends to use lines in places with a smaller mean
temperature.

\subsection{The relation between dark matter and baryons}

The baryons experience pressure forces, so their distribution on small
scales may not be the same as that of the dark matter. In our
technique, we model this effect by smoothing the dark matter
distribution to get the gas density using a 3D filter with a constant
filtering scale across the box ($k_f$). In analytic models, $k_f$
depends on the temperature of the IGM and on its reionization history.
The smoothing scale is one of the parameters in the grid so we
marginalize over it when obtaining constraints on other parameters.

\begin{figure}[tb]
\begin{center}
\leavevmode
\epsfxsize=9cm \epsfbox{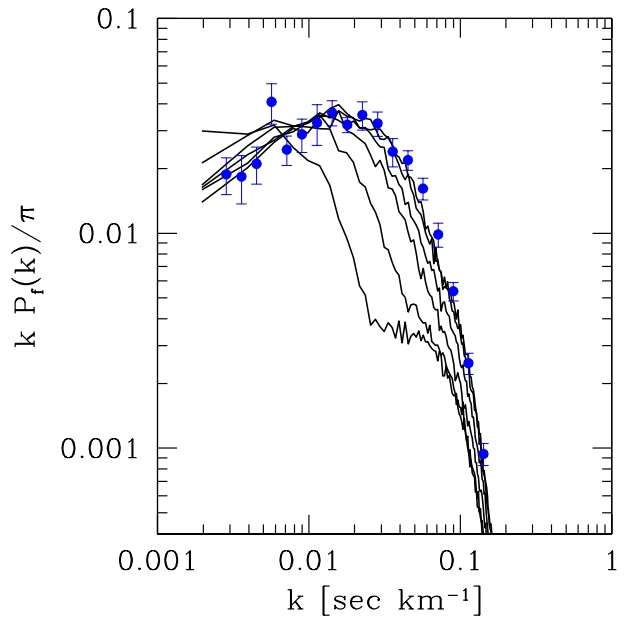}
\end{center}
\caption{Transmission power spectrum at $z=3$ for models with varying 
$k_f$ ($k_f=(5,10,20,30,40) {\rm h Mpc^{-1}}$ from bottom to top). All
other parameters are left constant, $\p=(.24, .7 ,
k_f, 400, .2,.7)$  . The data points are
from McDonald et al. (1999).}
\label{movie_kf}
\end{figure}

Even though we have $k_f$ as a free parameter to marginalize over, we
still get constraints on the temperature of the IGM.  It is
interesting to understand why that is the case, because both effects are a form
of smoothing. The key is that the $k_f$ smoothing of the density field
is done in 3D. After smoothing, a nonlinear transformation is
applied to the density to obtain the flux. This transformation shifts
power between scales, distorting the shape of the transmission power
spectrum. Figure \ref{movie_kf} illustrates the effect. For large
smoothing scales (small $k_f$), the shape of the transmission power spectrum is
totally wrong. 
Only when the smoothing by $k_f$ is subdominant
to smoothing produced by thermal broadening are the fits acceptable. 
Note that some of our higher $k_f$'s probably approach the resolution limit
of our simulations. We obtain lower limits on $k_f$,
$k_f > 10,30,25 \ \ {\rm h \, Mpc^{-1}}$ for $z=2.4,3,3.9$ which can be
compared with the Nyquist cut-off of the simulation $k_{Nyq}=50 {\rm h
Mpc^{-1}}$. Thus some care is needed when interpreting our lower
limits, as they may be somewhat sensitive to our resolution. In any
case, the conclusion that the 3D smoothing should be subdominant
compared with what is produced by the temperature should be robust
against increase in numerical resolution.

\subsection{The power spectrum of mass fluctuations}

\begin{figure}[tb]
\begin{center}
\leavevmode
\epsfxsize=9cm \epsfbox{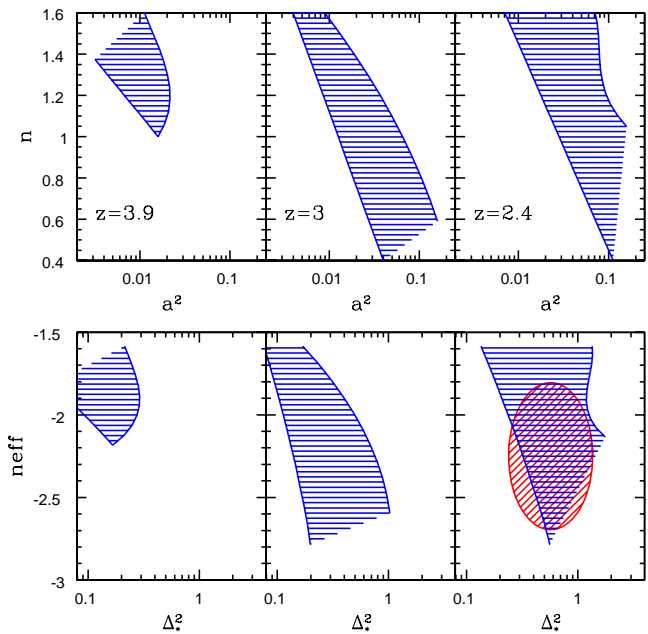}
\end{center}
\caption{The Constraints on amplitude and spectral index of the power
spectrum of initial fluctuations. In the top we show constraints in
our grid variables. On the bottom we have transformed this constraints
to amplitude and spectral index of the initial power spectrum at
$k=0.008 \ {\rm km^{-1}\ sec}$ to compare with the constraints form
\cite{croft99} shown with an ellipse (lines at $45^o$ angle).}
\label{sgnsconst}
\end{figure}

Finally we want to consider constraints on the power spectrum of mass
fluctuations. Figure \ref{sgnsconst} shows our constraints in the
amplitude/spectral index plane. The top panel shows directly the
contours in our primary variables $(a,n)$, the expansion factor of our
simulation and the primordial spectral index. 

In the bottom panel of figure \ref{sgnsconst}, we have changed our
contours to the $(\Delta^2, n_{eff})$ plane, where $\Delta$ is
the amplitude and $n_{eff}$ is the spectral index of the linear power spectrum
at $k=0.008 \ {\rm km^{-1}\ sec}$.  We did this to compare with the
constraints from Croft et al. (1999) shown with a red ellipse.  Although we
agree in the range of allowed amplitudes, we disagree in the ability
of the data to constrain the spectral index. The reason we cannot
constrain the spectral index from the shape of the transmission power
spectrum can be traced to our having $k_f$ as a free parameter. Figure
\ref{movie_kf} shows that changes in $k_f$ modify the low $k$ slope of
the transmission power spectrum.

\begin{figure}[tb]
\begin{center}
\leavevmode
\epsfxsize=9cm \epsfbox{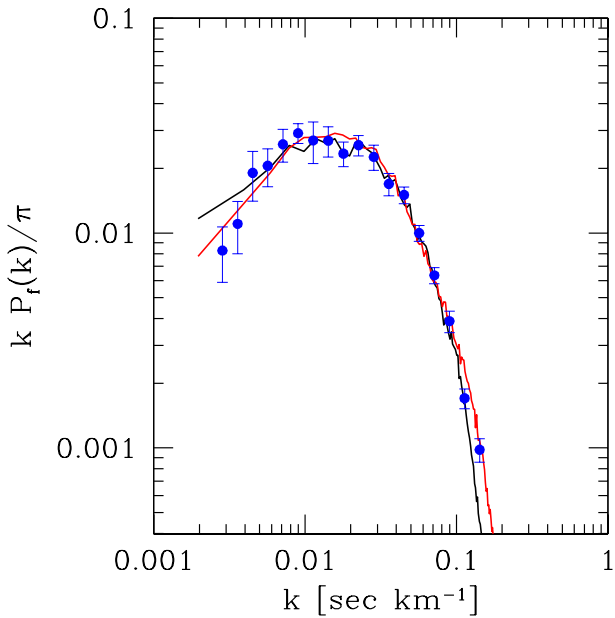}
\end{center}
\caption{Two models with very different spectral index that have
similar flux power spectra together with the $z=2.4$ data. One model
has $\p=(.41, .4 , 70, 350, .3, .82)$ with $\chi^2=11$ and the other
has $\p=(.14, 1.6, 20, 250, .2, .82)$ with $\chi^2=10$.}
\label{bestsvns}
\end{figure}

In figure \ref{bestsvns} we plot two models with very similar transmission
power spectrum but with $n=0.4$ and $n=1.6$. The difference in slope,
which should otherwise be noticeable at low $k$, is compensated by a
change in $k_f$. This change in $k_f$ should produce changes in the power
spectrum at high $k$, which is offset by a change in $T_0$.
Thus the power spectrum
data alone cannot accurately constrain the shape of the matter power
spectrum unless other information is used to constrain $k_f$.

\section{Improvements for the future}\label{future}

The objective of this work was to show that detailed analysis of the
transmission power spectrum can be used to extract useful information
about the cosmological model and the properties of the IGM at high
redshift. In order to make progress and extract more information from
the data, there are several aspects of our work that could be improved
in the future.

From an observational perspective, we need to improve the determination
of the transmission power spectrum. Important ingredients will be to obtain a
reliable covariance matrix for the different band powers and to
determine the window functions that relate the observed power with the
theoretical predictions. It would also be desirable to measure the power
spectrum of $\delta F / \bar F$ after trend-removal rather than 
the power spectrum of $\delta F$ after continuum-fitting (Hui et al. 2000), 
to minimize systematic errors due to continuum placement.

From a theoretical perspective, one should study in more detail the
relation between baryons and dark matter. Hydrodynamic simulations as
well as the hydro-PM approximation (\cite{gh98}) should be used to understand and
quantify the possible biases introduced when extracting different
parameters from PM simulations. Our method requires exploring a large number
of models, so it will rely on dark matter only simulations (or the
hydro-PM variant) for the
foreseeable future. 

A very interesting possibility is to try to determine if there are
spatial fluctuations in the equation of state of the IGM. This will require
measuring the transmission power spectrum and inferring a temperature
for individual portions of spectra along different lines of sight and
comparing results. A comparison will only be meaningful if one
understands the expected distribution for these band-powers from
simulations. It is important to know what one expects from cosmic
variance alone and how different band powers are correlated in models
with uniform equation of state. 

Another direction which should be pursued is the combination of the
power spectrum analysis with other information, in particular
the probability distribution
function (PDF) of the transmission. It will be interesting to know if
the PDF can break some of the degeneracies we encountered. It is
expected however that some degeneracies between parameters 
will remain (Theuns et al. 2000).

\section{Conclusions}\label{conclusions}

We have implemented a simple likelihood method to constrain the
parameters of a Lyman $\alpha$ forest model from transmission 
power spectrum data.  We
used a 6 parameter grid and a marginalization method based on previous
work for CMB data Tegmark \& Zaldarriaga (2000). It is important that we fit for the 6
parameters simultaneously, as we find several parameter degeneracies
that prevent us from extracting tighter constraints.

Perhaps our most interesting constraint is that on the equation of
state of the gas. Our results are in good agreement with those of
McDonald et al. (2000). We find a temperature that is somewhat higher than
might have been expected and see little evidence for strong evolution of
the temperature with redshift. Our determinations of the mean temperature
and of temperature-density slope are highly degenerate. 

As for the constraints on the mass power spectrum, we
find amplitude constraints that are in agreement with those found
by Croft et al. (1999). On the other hand, we are unable to place
accurate constraints on the spectral index of the power spectrum, mainly due to
an uncertainty in the relation between baryons and dark matter,
parametrized by a smoothing scale $k_f$. 
This uncertainty is in part a consequence of our using dark
matter only simulations to model the forest. Thus if one could
construct a grid using hydrodynamic simulations, perhaps using the 
hydro-PM method of Gnedin \& Hui (1998), some of this uncertainty could
be removed. However, that part of the uncertainty in $k_f$ which is
due to our ignorance of the reionization history will remain.

An important finding regarding smoothing is that the 3D smoothing
parametrized by $k_f$ is less important than the 1D smoothing due to
thermal broadening in determining the small scale transmission power
spectrum. This is what allows us to determine the temperature of the
IGM from the rapid fall-off in the small scale power. A $k_f$ small
enough to affect the fall-off would also modify the shape of the
transmission power spectrum sufficiently to make it an unacceptable
match to observations.

{\bf Acknowledgements} We would like to thank Joop Schaye for useful
discussions.  M.Z. is supported by the Hubble Fellowship
HF-01116-01-98A from STScI, operated by AURA, Inc. under NASA contract
NAS5-26555.  L.H. is supported by NASA NAG5-7047 and the Taplin
Fellowship at the IAS, and by an Outstanding Junior Investigator Award
from the DOE. MT is supported by NASA grant NAG5-9194 and NSF grant
AST00-71213 and the University of Pennsylvania Research Foundation.

\end{document}